# Characterizing Voltage Contrast in Photoelectron Emission Microscopy

Vinod K. Sangwan[1,2], Vincent W. Ballarotto[2], Karen Siegrist[3] and Ellen D. Williams[1,2]

[1] Department of Physics, University of Maryland, College Park, Maryland 20742

[2] The Laboratory for Physical Sciences, College Park, Maryland 20740

[3] Applied Physics Laboratory, Laurel, Maryland 20723

**Abstract:**

A non-destructive technique for obtaining voltage contrast information with photoelectron emission microscopy (PEEM) is described. Samples consisting of electrically isolated metal lines were used to quantify voltage contrast in PEEM. The voltage contrast behavior is characterized by comparing measured voltage contrast with calculated voltage contrast from two electrostatic models. Measured voltage contrast was found to agree closely with the calculated voltage contrast, demonstrating that voltage contrast in PEEM can be used to probe local voltage information in microelectronic devices in a non-intrusive fashion.

**Introduction:**

Using voltage-induced image contrast for analysis of integrated circuits (IC) is a well known technique [1,2]. Both the scanning electron microscope (SEM) and focused-ion beam tools use this type of contrast to test circuits on IC chips. This is done by applying an external bias to a specimen and analyzing variations in image contrast to determine if a circuit functions properly. However, image as well as specimen degradation can occur from prolonged exposure to a charged-particle beam, and thus can limit device testing [3]. Limitations are mainly due to deleterious beam effects such as inelastic scattering of electrons, inadvertent ion implantation, as well as surface sputtering, all of which can permanently damage the device under test.

Imaging surfaces with low energy electrons with techniques like low energy electron microscopy (LEEM) [4-8] and photoelectron emission microscopy (PEEM) [9-12] is also well established. These techniques generate real time in-situ images of surface that can be used to analyze the device. In PEEM the imaging electrons are generated by near-threshold photoemission, and thus the image intensity is sensitive to surface fields. This means that non-destructive voltage contrast can be acquired with PEEM. However, the lateral fields responsible for the contrast can arise from either an externally applied bias or in combination with surface topography. By varying the external bias, voltage contrast and topography contrast can be readily distinguished.

In this paper, we describe how PEEM voltage contrast arises when an external voltage is applied to a device. To achieve this we have fabricated test structures that allow the two types of image contrast to be distinguished. The voltage contrast behavior is characterized by comparing measured voltage contrast with calculated voltage contrast from two different models. In the first model, electron trajectories are numerically simulated to quantify the intensity profiles. In the second model, an electron optics formulation developed by Nepijko [13-16] is used to quantify the intensity profile. Agreement between the measured and calculated voltage contrast suggests that simple electrostatic models are sufficient to describe voltage contrast in PEEM. Although fundamentally different, PEEM voltage contrast can be used in the same manner as voltage contrast is used in SEM and FIB tools.

**Experimental:**

To generate potential gradients at the sample surface in a controlled manner, layered device structures were fabricated where the top metal layer is electrically isolated from the bottom metal layer. This process was used previously for fabrication of devices used to characterize topography-induced contrast in PEEM [17]. Thin films of Ti (500 nm), $SiO_2$ (200 nm) and Ti (100 nm) were successively deposited on a clean Si wafer by electron-beam evaporation. The tri-layered wafers were then patterned by photolithography using a positive photoresist (908 20 HC, Fujifilm). The photoresist was used as an etch mask and the exposed surface was etched to the bottom Ti layer in a reactive ion etcher with a $SF_6$ plasma (pressure = 525 mTorr, rf power = 550 W). The etching time was carefully controlled in order not to overexpose the underlying Ti thin film. Finally, the photoresist was removed in heated n-methyl pyrolidinone to obtain sets of isolated raised lines which could be independently biased. Thickness and profile of the resulting Ti and $SiO_2$ bi-layer was characterized by profilometry and SEM. A cross-sectional schematic of the sample is shown in Fig. 1(a). A scanning electron micrograph of the sample (see Fig. 1(b)) confirms the desired vertical profile of the bi-layer with a total thickness of 350 nm.

PEEM imaging of the devices was carried out in a horizontally configured microscope (Staib) [18,19]. The samples were illuminated at 70º from the surface normal by a 100 W Hg short-arc lamp, which has a high energy cut-off of approximately 5.1 eV. Samples were transferred into a UHV ($5 \times 10^{-9}$ to $10^{-8}$ Torr) chamber through a rapid entry load-lock system. A computer controlled 5-axis positioning stage was used to carefully position the sample approximately 5 mm from the aperture lens. The images were recorded using a 16-bit CCD camera which produces a 1280 x 1024 bitmap and stored in 12-bit tif format. All the images for voltage contrast were taken with an exposure time of 0.8 sec and averaged over 16 exposures. The image intensity was quantified by averaging line-scans (using IDL 6.0) along 200 parallel lines to obtain an averaged intensity profile. To correct for background intensity fluctuations, a 10$^{th}$ order polynomial was first fitted to the average background intensity profile while substituting the edge intensity curves with linear interpolation, and then subtracted from the original line scan data. Image contrast can be characterized by any of three parameters: the area under the curve $A$, full-width half-minimum and depth of the intensity dip $d$. However $d$ and tend to be more sensitive to the fluctuations in the intensity peak than $A$. Moreover, $A$ is also more amenable in comparing with numerical simulations. Therefore, we quantified the experimental contrast using the area $A_{expt}$, which is calculated as the area under the intensity profile between the two points where intensity increases by 90 % of the minimum intensity at the dip.

To illustrate voltage contrast we briefly describe a typical PEEM image of the device in Fig. 2(a). Shown in the image is a pair of 16 μm wide structures with a separation of 16 μm. The top Ti layer in the Ti/SiO$_2$/Ti structure at the left (1) is shorted to the underlying Ti surface and grounded in order to get direct comparison between the biased and unbiased structures. The Ti line on the right (2) is externally biased at 5 V with respect to the background Ti surface. The contrast observed along the edge of Ti line 1 is solely due to topography, whereas the contrast observed along the edge of Ti line 2 is due to both topography and the external bias. Fig. 2(b) shows an intensity profile averaged from 200 line scans taken along a group of dashed lines shown in Fig. 2(a). The 4 features arise from the contrast at the edges of the lines. In this particular case, $A$, and $d$ for the biased (5 V) structure are approximately 6.5, 2.5 and 2 times, respectively, larger than the corresponding parameters for the unbiased structure. The external bias increases contrast at the edges of line 2 in magnitude and lateral extent. Therefore calculating the electric field near the edge of the metal line is necessary to determine the effect on image intensity profile.

**Numerical Model:**

The numerical model used to determine the intensity profile calculates the electron trajectories in the PEEM immersion lens. The calculation was done by SIMION 7 [20] which uses a finite difference method to calculate the electric potential, and a Runga-Kutta method to calculate the electron trajectories. The top plate of the simulation box (100 μm x 100 μm; grid size: 5000 x 5000) was biased at 200 V (10kV×100μm /5mm) to obtain the desired $E_0$ (2 V/μm). Neumann boundary conditions were set on the side edges. The initial kinetic energy of the electrons was assumed to be 0.3 eV. The electrons were emitted at angles -10°, -6°, -2°, 0°, 2°, 6°, 10° with respect to the surface normal at each of the 150 regularly spaced emission points on both sides of the edge of the sample. A sample of electron trajectories from 12 emission points near the sample edge is shown in Fig. 3(a). The top metal layer of the sample was biased at -1 V. The position and velocity of the electrons at the top plate of simulation box were used as initial conditions to simulate the electron trajectories in the bigger simulation space (from $y = 100$ μm to $y = 5$ mm i.e. whole PEEM immersion lens). The field perturbation due to the sample is neglected in this region and the analytical expression of electron trajectories in a uniform electric field $E_0$ is used to trace the parabolic rays of the electrons (using MATLAB). The array of $x$ coordinates of electrons at the aperture lens plane ($y = 5$ mm) is taken to represent the distribution of the electrons exiting the acceleration region of the microscope.

The density of exiting electrons $d_0$ is calculated as a function of $x$ by counting the number of electrons in uniformly spaced segments along the $x$ coordinate. Due to the finite numbers of electron rays in the simulation, $d_0(x)$ showed fluctuation between discrete levels. A smoother distribution $d_1(x)$ was obtained by doing a running average over five nearest neighbors in $d_0(x)$. This process was repeated again in $d_1(x)$ to obtain an even smoother distribution $d_2(x)$. Fig. 3(b) illustrates how this averaging technique minimizes the numerical noise generated in the calculation. The numerical electron density distribution can be thought of as a result of the displacement of electrons from the intensity dip to the intensity peak due to the lateral component of the electric field. However in practice we have a contrast aperture (diameter 70 nm) in the back focal plane of the microscope and most of the electrons deviated by large angles are eliminated before reaching the imaging plane. Therefore, we do not observe the bright maximum in the PEEM images here, as suggested by the calculation. To quantify the numerical contrast we ignore the maximum peak in intensity profile and only consider the area of the minimum peak $A_{num}$ in the same manner as was done for the measured contrast (i.e. the area between the two points where intensity increases by 90 % of the minimum intensity at the dip).

**Analytical Model:**

We also utilized a analytical treatment [13-16,21] to quantitatively predict the electron density distribution in the PEEM image due to the local electric fields on the sample surface. This method has been used earlier to model image deterioration in PEEM [16] and surface potential mapping of microelectronic devices in PEEM [15]. We use it here for modeling PEEM contrast induced by externally biasing a specimen. Deflection of the electron trajectories due to the local fields is calculated to first order approximation. A Cartesian coordinate system is chosen such that the *x-y* plane coincides with the sample surface, and the *z*-axis coincides with optical axis of the cathode lens in PEEM. The long rectangular samples studied here allow us to assume that the local electric potential varies only along one of the coordinates (chosen as *x* in Fig. 2). We also assume absence of the contrast aperture in the model used here. The photoelectrons are traced through the acceleration region without any restriction, and thus the total electron current is conserved. The shift in electron trajectories $S$ as a function of local electric potential $\phi$ is derived from Dyukov et al. [13] and is given as [21]

$$S(x) = \frac{1}{\pi E_0} \int_{-\infty}^{\infty} \phi'(x-\xi) \ln(1+\frac{l^2}{\xi^2}) d\xi, \qquad (1)$$

where $E_0$ is accelerating field, $l$ is the distance between the sample surface and aperture lens. When the top layer of the device is shorted to the grounded background surface, the surface of the sample is an equipotential. Thus, in the vicinity of the surface the electric potential can be assumed to follow the contour of the topography (i.e. $\phi_{topo}(x) = E_0 h(x)$). Furthermore, the rectilinear step can be approximated as a smoothly varying step of height $h_0$ and half-width $a$, $h(x) = \frac{h_0}{\pi} \tan^{-1} \frac{x}{a}$. When the surface of the metal line is biased with respect to the grounded background surface, the surface of the metal line is not an equipotential. However, the effect of external bias on the topography can be taken into account by changing the height of the step such that it becomes an equipotential surface. An external bias of voltage $V_{bias}$ would correspond to the change in height of the step by $V_{bias}/E_0$. Therefore the local electric potential for a biased sample takes the form

$$\phi_{bias}(x) = \frac{(E_0 h_0 + V_{bias})}{\pi} \tan^{-1} \frac{x}{a}. \qquad (2)$$

Differentiation and substitution of this expression for the electric potential in Eq. 1 yields a shift of the form

$$S(x) = -\frac{(E_0 h_0 + V_{bias})}{2\pi E_0} \ln \frac{x^2 + (1+a)^2}{x^2 + a^2}. \qquad (3)$$

The role of the local electric field is to redistribute the electron current density in a plane as dictated by the expression of shift given above. For a planar sample, the electrons are nominally collected at a point $x$. However, in the presence of a local perturbation due to external biasing or topography, the electrons are collected at a point $x + S(x)$. The redistributed electron density is expressed in terms of the shift as [14]

$$j(x+S) = \frac{j_0(x)}{1 + dS/dx} \qquad (4)$$

where $j_0(x)$ is the distribution of electrons in case of a planar sample and is a constant (designated $j_0$). After substituting the expression for $S(x)$ in equation (4) we get

$$j(x+S) = j_0 \left( 1 - \frac{(E_0 h_0 + V_{bias})(a^2 + x^2)\left(\frac{2x}{a^2+x^2} - \frac{2x((a+l)^2 + x^2)}{(a^2+x^2)^2}\right)}{2E_0\pi((a+l)^2 + x^2)} \right)^{-1}$$

(5)

where we can assume that $j_0 = 1$ since the intensity is in arbitrary units. The expression for $j(x)$ in Eq. (5) gives the intensity profile for a sample with a specified height $h_0$, applied voltage $V_{bias}$. Other parameters were kept the same throughout the calculations reported here ($E_0 = 2$ V/μm, $l = 5$ mm and the edge smoothness parameter $a = 100$ nm). Since this analytical formulation does not take into account the contrast aperture, a peak in the intensity profile is produced next to the dip. The analytical contrast is characterized as the area under the intensity dip $A_{theo}$ in the same manner as the numerical contrast $A_{num}$ and measured contrast $A_{expt}$.

**Results and Discussion:**

First we consider a set of PEEM images of a 300 nm tall isolated Ti line that were acquired as the external voltage was varied from – 6 V to 6 V in steps of 1 V. The inset in Fig. 4 shows a set of corrected line scans taken over the edge of the line for 0 V to -5 V, and clearly illustrates the effect of biasing the line. As the bias increases (both polarities), the image intensity near the edge of the line decreases and the lateral extent of the affected area increases. Both of these features can be seen by examining $A$ and the corresponding $\Gamma$ as shown in Fig. 4, which also illustrates that either $A$ and $\Gamma$ can be used to quantify voltage contrast. In addition, the measurements show that both quantities have a minimum near $V_{bias}$ = 1 V, not at $V_{bias}$ = 0 V. As shown below, magnitude of $A$ at $V_{bias}$ = 0 V is due to the height of the metal. Although not shown here, the measured intensity starts to saturate at higher biases ($|V_{bias}| > 8$ V) due to the limited dynamic range of the channel plate.

To model the observed behavior, numerical and analytical calculations of intensity profiles were done for a sample defined as lying in the region $x > 0$ with the edge of interest at $x = 0$. For the numerical model the electron density distribution $d_2(x)$ was calculated for biases ranging from -7 V to 7 V in 0.5 V increments. For the analytical model the redistributed electron density $j(x + S)$ was calculated for biases ranging from -8 V to 8 V in steps of 0.2 V. Results for $d_2(x)$ and $j(x + S)$ are shown in Fig. 5(a) and Fig. 5(b), respectively, for -3 V to 1 V. Notice that for $x > 0$ both $d_2(x)$ and $j(x + S)$ decreased as the bias decreased to -3 V. $d_2(x)$ is minimized around $V_{bias}$ = 0.5 V and $j(x + S)$ is minimized at $V_{bias} = 0.6$ V (where $j(x + S)$ becomes constant and equal to 1). This behavior is similar to the measured intensity profile. Also, we find that further increase in the bias inverts the calculated electron densities (e.g. black line in Fig. 5(a) and 5(b) for $V_{bias} = 1$ V) because the local electric field changes direction.

For direct comparison with the measured intensity profiles, the calculated electron densities were scaled such that $A_{expt} = A_{num} = A_{theo}$ for $V_{bias} = -1$ V. In the numerical case, the scaling factor corresponds to the ratio of the actual density of photoelectrons emitted from the surface to the density of electrons assumed in the calculations. In the analytical case, the scaling is equivalent to setting the parameter $j_0$ a value other than 1 in Eq. 5. These scaling factors are kept constant in all the voltage contrast results reported here. However we note that the relative contrasts do not change significantly by changing the value of $V_{bias}$ used to scale the data. The measured intensity profile at $V_{bias} = -1$ V is compared with the scaled $d_2(x)$ (blue circles) and $j(x + S)$ (red squares) in Fig. 6. Neglecting the peaks outside the sample region, we find that there is good agreement between the scaled calculated electron densities and the measured intensity profile.

Finally, the voltage contrast was quantified by extracting the area under the curves for the calculated electron distributions and comparing with the areas extracted from the measured intensity profiles. As shown in Fig. 7, there is excellent agreement between the measured and calculated voltage contrast as a function of applied bias. We find that the minimum numerically calculated voltage contrast occurs at $V_{bias} = 0.5$ V, which is near the minimum voltage contrast measured at $V_{bias} = 1$ V. The minimum analytically calculated contrast occurs at $V_{bias} = 0.6$ V and is equal to zero. The minimum voltage contrast should occur when the lateral component of the externally applied field is roughly equal and opposite the lateral component due to the height of the line. For a line $h_0$ = 300 nm tall in a uniform electric field this corresponds to a $V_{bias}$ = 0.6 V ($V_H = |\vec{E_0}| \times h_0 = 10\text{kV} \times 300\text{nm}/5\text{mm}$). Minimizing (or eliminating) the contrast at the edge of the metal line can, in principle, allow for determination of the height of the structure.

More careful modeling of the external electric field near the edge of the metal line shows that $E_x$ can become significantly larger than the accelerating field. A finite element method (Poisson Superfish [22]) was used to calculate the electric field near the line edge as the external bias was varied from - 8 V to 8 V in 0.5 V increments. In Fig. 8, $E_x$ at 30 nm above the metal line ($1.1 \times h_0$, $h_0$ is height of the sample) is plotted as a function of $x$ for several sample biases. The calculation shows that $E_x$ is much less than 1 V/μm except within 1 μm of the line edge, where it increases significantly. Near the step edge, $E_x$ can become as large as 15 V/μm, approximately 7 times larger than $E_0$ used to image our devices. This means that the applied field can no longer be considered a perturbation to the accelerating field as it is significantly larger. Instead, the electrons traveling in this physical area will be significantly altered from their characteristic curves. Therefore voltage-induced contrast appears only in regions where the externally applied field is sufficient enough to affect PEEM image intensity (i.e. near the edges of the metal lines). This is different than voltage contrast in SEM (or FIB) where the image intensity of the whole structure is affected by the applied bias. $E_x$ is observed to be zero everywhere for $V_{bias}$= 0.6 V where minimum contrast was also observed (red curve in Fig. 8).

As we have shown here, PEEM voltage contrast can be generated by applying either a positive or negative bias to a metal line. Observing the change in the edge contrast of metal line in PEEM will thus allow identification of lines that are continuous with a voltage probe. Conversely, a metal line that is not continuous would not produce changes in image contrast. Therefore, this technique can be used to test IC functionality in the same manner as SEM and FIB tools. In applications where rapid inspection of large areas is desired, the large field of view and high information throughput of PEEM, which is not a scanning technique, may be preferable.

**Conclusion:**

We have demonstrated that voltage contrast in PEEM can be readily achieved and distinguished from topography contrast. Using simple devices with isolated metal lines, the measured voltage contrast agrees well with the calculated voltage contrast that is based on electrostatic effects to the distribution of imaging electrons. These models show that the electron distribution leaving the accelerating region of the microscope is affected by local electric fields on the specimen. Since the electron distribution can be influenced near the edge of a biased structure, the corresponding image intensity can be systematically varied. As a result, voltage contrast information can be obtained with PEEM. Thus, it is straightforward to distinguish between an open and a break in a metal line by simply tuning the external bias and observing the edge contrast response. An interesting corollary is that the height of a structure can be measured by determining the bias necessary to minimize the voltage contrast. PEEM voltage contrast offers a useful method to probe local potential variations in electronic devices.

**Acknowledgements:**

This work was supported by the Laboratory for Physical Sciences.

**Figure captions:**

Figure 1. (a) Schematic of a tri-layered sample for voltage contrast. The top layer can be independently biased with respect to the background Ti surface. (b) An oblique SEM micrograph of the sample shows different layers.

Figure 2. a) A PEEM image of a sample showing the effect of bias on imaging contrast. b) Intensity scans were averaged over 200 lines in the region depicted by dotted lines in (a). Intensity dips in the biased sample show increase in full width half minimum ( ), depth (*d*), and hence the area (*A*) under the curves.

Figure 3. a) Electron trajectories simulated using SIMION 7 when the sample is biased at -1 V. The electrons emitted at near the sample edge with initial angles -10°, -6°, -2°, 0°, 2°, 6°, 10° with respect to the surface normal. b) Digitized density of electrons $d_0(x)$ at the aperture lens plane and averaged densities $d_1(x)$ and $d_2(x)$ plotted as a function of position, *x*. The *y* axis corresponds to $d_0(x)$ and subsequent plots are shifted by 3 units.

Figure 4. Area under the curve *A* and the full width half minimum    as function of bias on the top Ti line of the sample with height of 300 nm. The inset shows intensity profile of the edge of the line for bias varying from 0 V to -5 V (as indicated by the arrow).

Figure 5. a) Electron distribution $d_2(x)$ obtained from the numerical method is plotted against the distance across the line edge for selected value of biases.   b) Electron

distribution $j(x + S)$ obtained from the analytical approach is plotted against the distance across the line edge for corresponding biases. The sequence of the intensity peak and the dip changes at $V_{bias}$ = 0.5 V. Part of curves under the shaded area is not considered for characterization of image contrast.

Figure 6. Intensity profiles obtained from two analyses (blue and red curves for numerical method and analytical method, respectively) are compared with intensity profile obtained from PEEM image. The peak in intensity (bright line) is absence in case of experimental curve due to the contrast aperture.

Figure 7. The area under the intensity dips from measured and calculated intensity profiles is plotted against the external bias of the sample. Since the three data sets are in arbitrary units they are scaled to be equal to 1 at $V_{bias}$ = -1V.

Figure 8. $E_x$ along the dotted line in inset at a height $h_1$ = 1.1 x $h_0$ (330 nm) is plotted as a function of $x$. Magnitude of $E_x$ is zero for all $x$ when the sample is biased at 0.6 V.

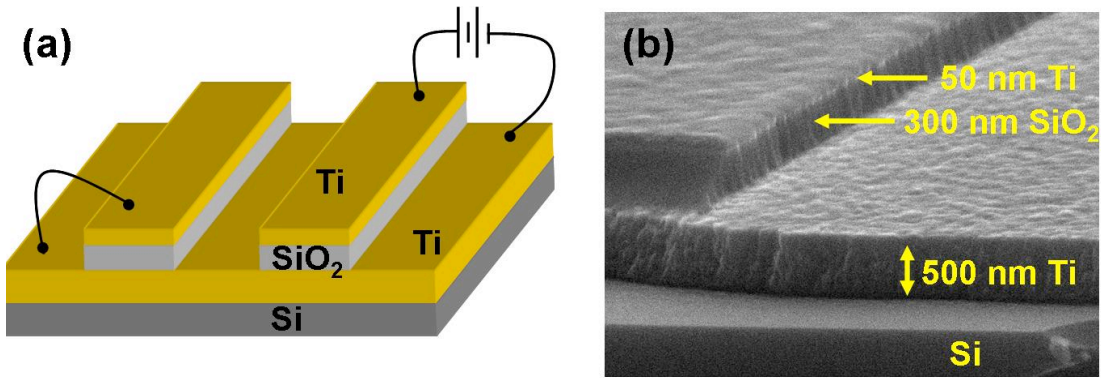

Figure 1. (a) Schematic of a tri-layered sample for voltage contrast. The top layer can be independently biased with respect to the background Ti surface. (b) An oblique SEM micrograph of the sample shows different layers.

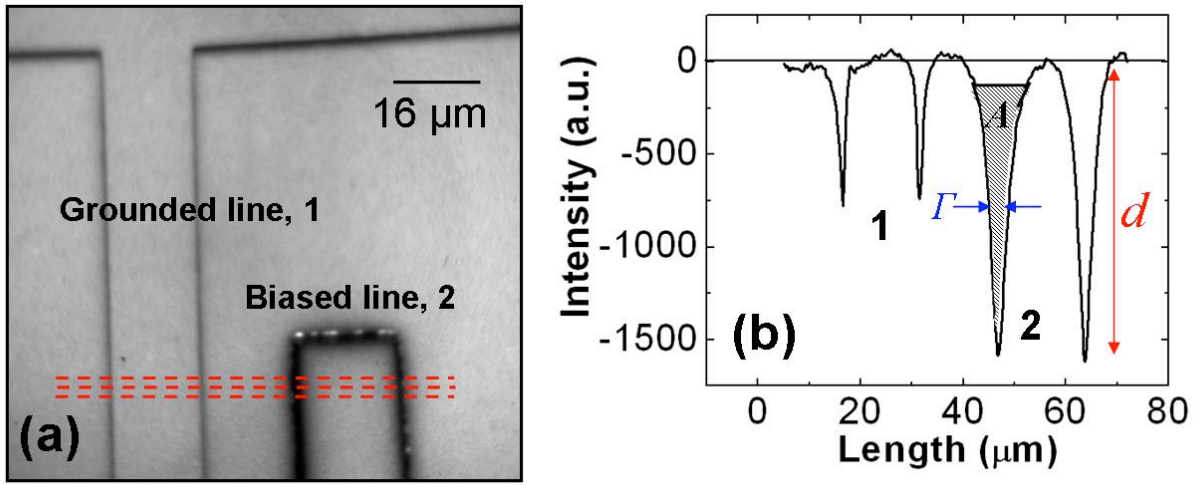

Figure 2. a) A PEEM image of a sample showing the effect of bias on imaging contrast. b) Intensity scans were averaged over 200 lines in the region depicted by dotted lines in (a). Intensity dips in the biased sample show increase in full width half minimum ( ), depth (*d*), and hence the area (*A*) under the curves.

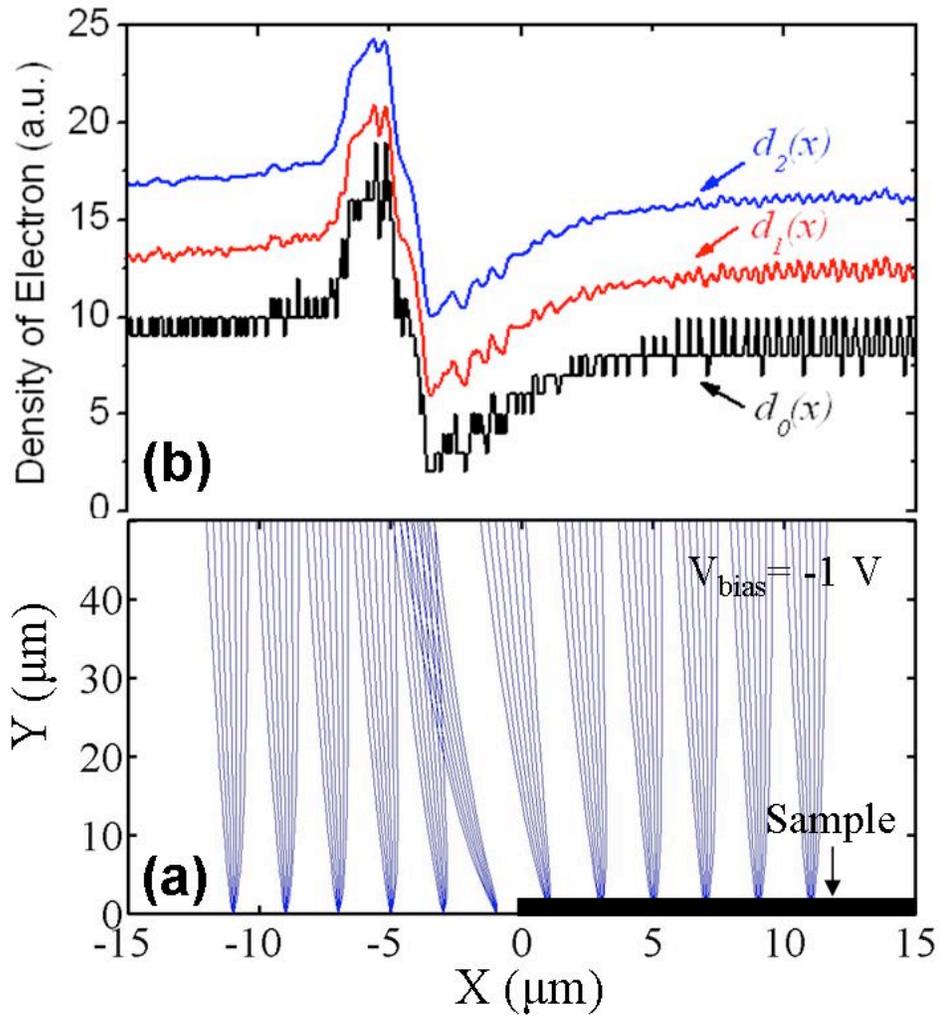

Figure 3. a) Electron trajectories simulated using SIMION 7 when the sample is biased at -1 V. The electrons emitted at near the sample edge with initial angles $-10°$, $-6°$, $-2°$, $0°$, $2°$, $6°$, $10°$ with respect to the surface normal. b) Digitized density of electrons $d_0(x)$ at the aperture lens plane and averaged densities $d_1(x)$ and $d_2(x)$ plotted as a function of position, $x$. The $y$ axis corresponds to $d_0(x)$ and subsequent plots are shifted by 3 units.

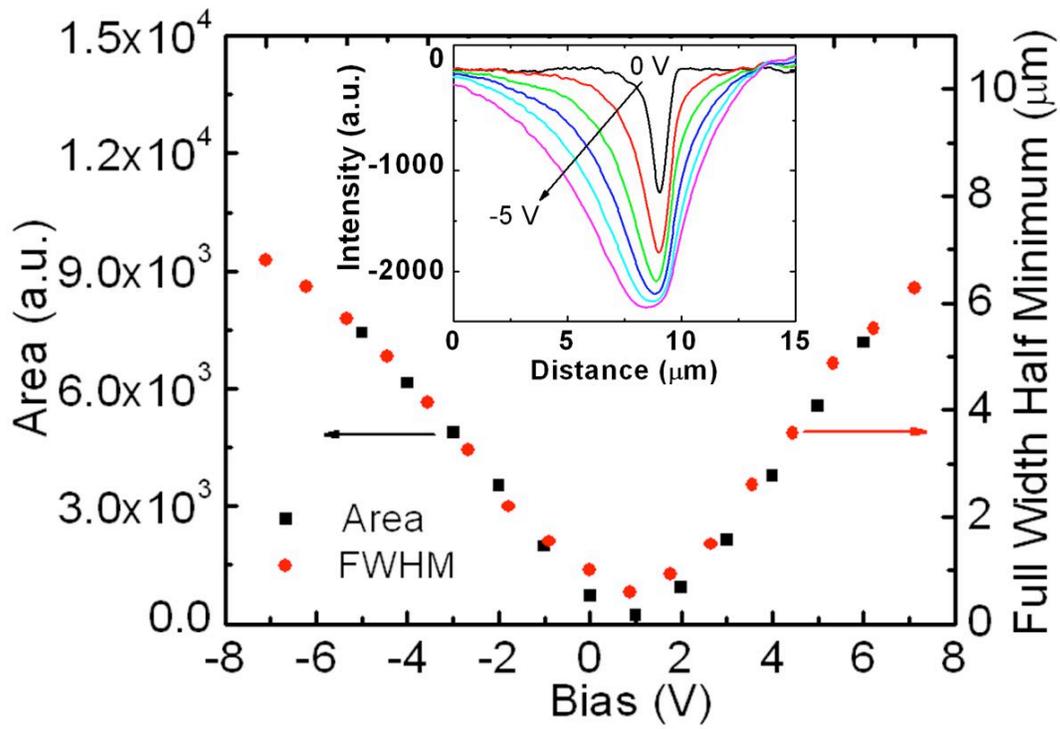

Figure 4. Area under the curve *A* and the full width half minimum as function of bias on the top Ti line of the sample with height of 300 nm. The inset shows intensity profile of the edge of the line for bias varying from 0 V to -5 V (as indicated by the arrow).

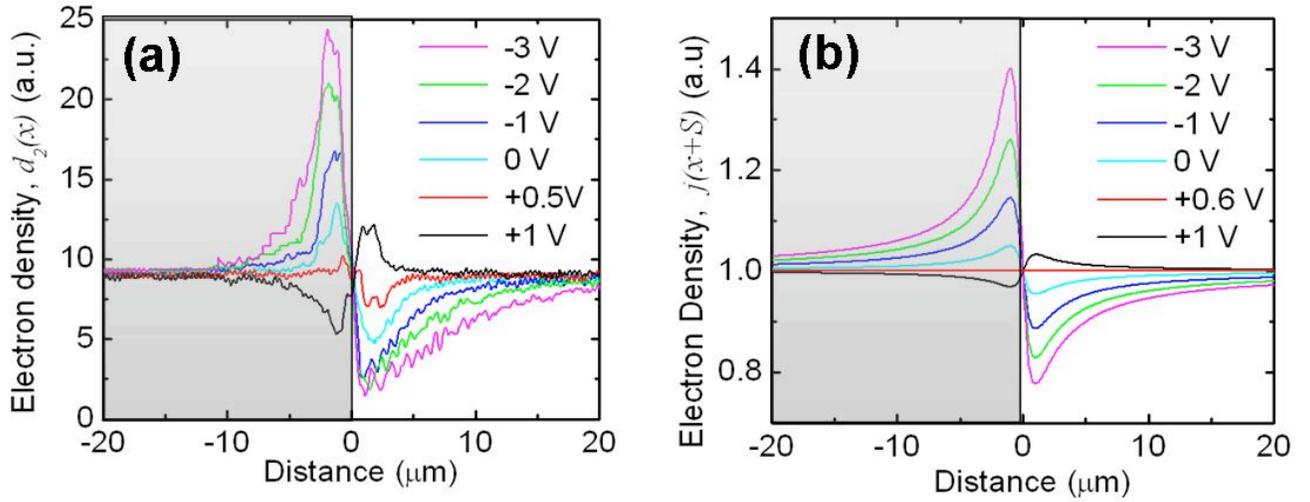

Figure 5. a) Electron distribution $d_2(x)$ obtained from the numerical method is plotted against the distance across the line edge for selected value of biases. b) Electron distribution $j(x + S)$ obtained from the analytical approach is plotted against the distance across the line edge for corresponding biases. The sequence of the intensity peak and the dip changes at $V_{bias}$ = 0.5 V. Part of curves under the shaded area is not considered for characterization of image contrast.

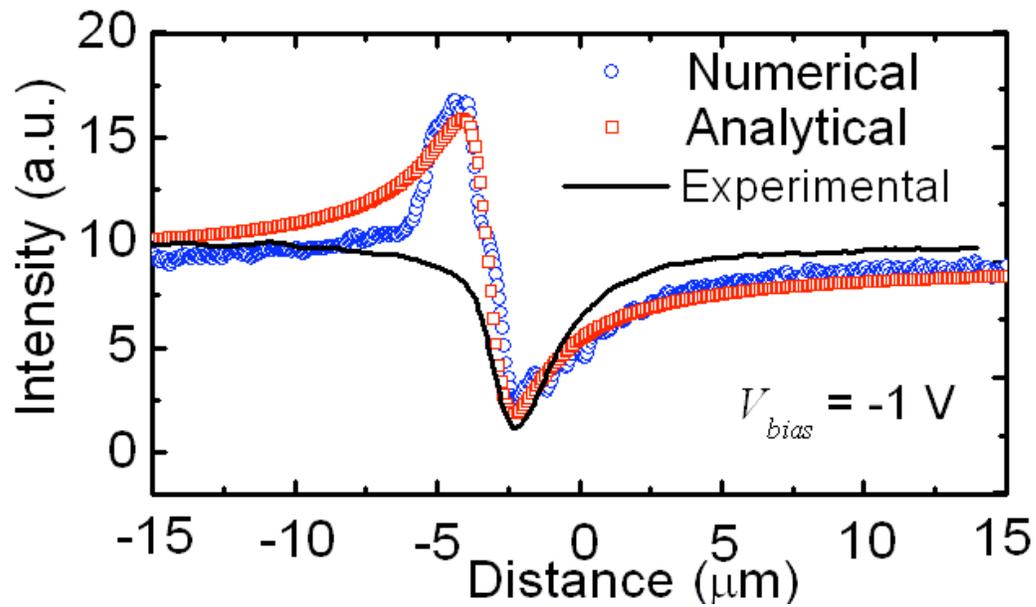

Figure 6. Intensity profiles obtained from two analyses (blue and red curves for numerical method and analytical method, respectively) are compared with intensity profile obtained from PEEM image. The peak in intensity (bright line) is absence in case of experimental curve due to the contrast aperture.

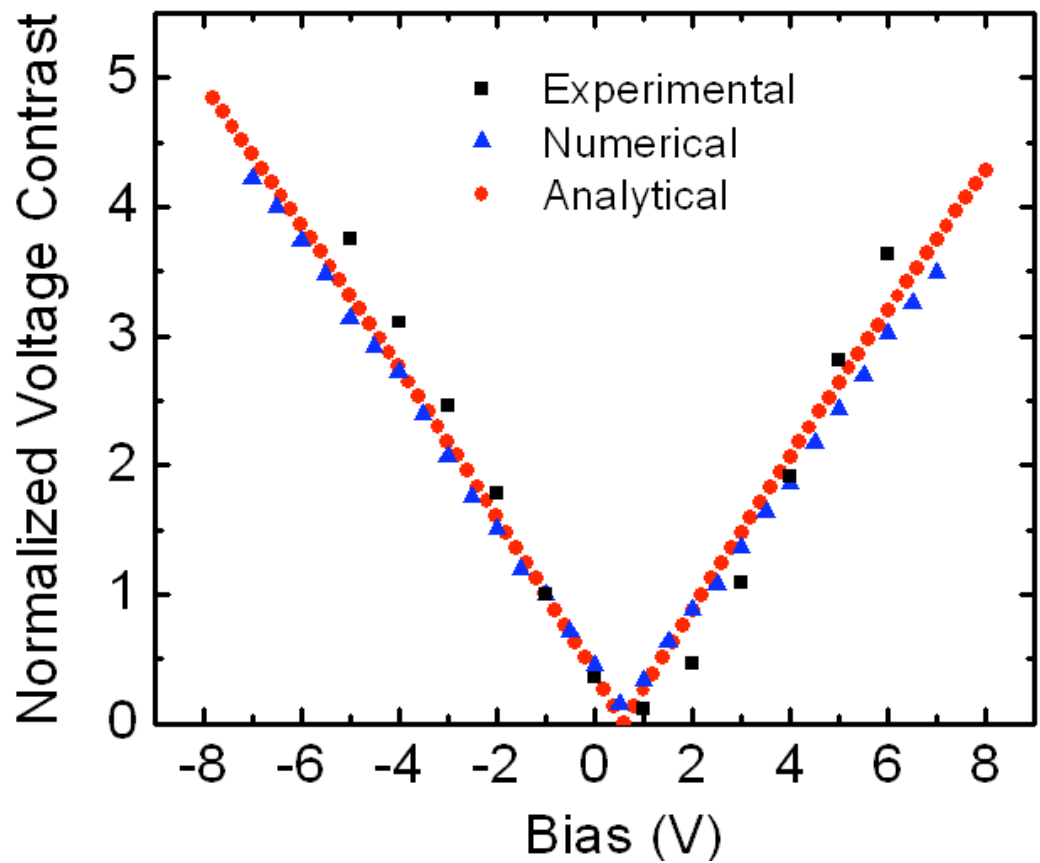

Figure 7. The area under the intensity dips from measured and calculated intensity profiles is plotted against the external bias of the sample. Since the three data sets are in arbitrary units they are scaled to be equal to 1 at $V_{bias}$ = -1V.

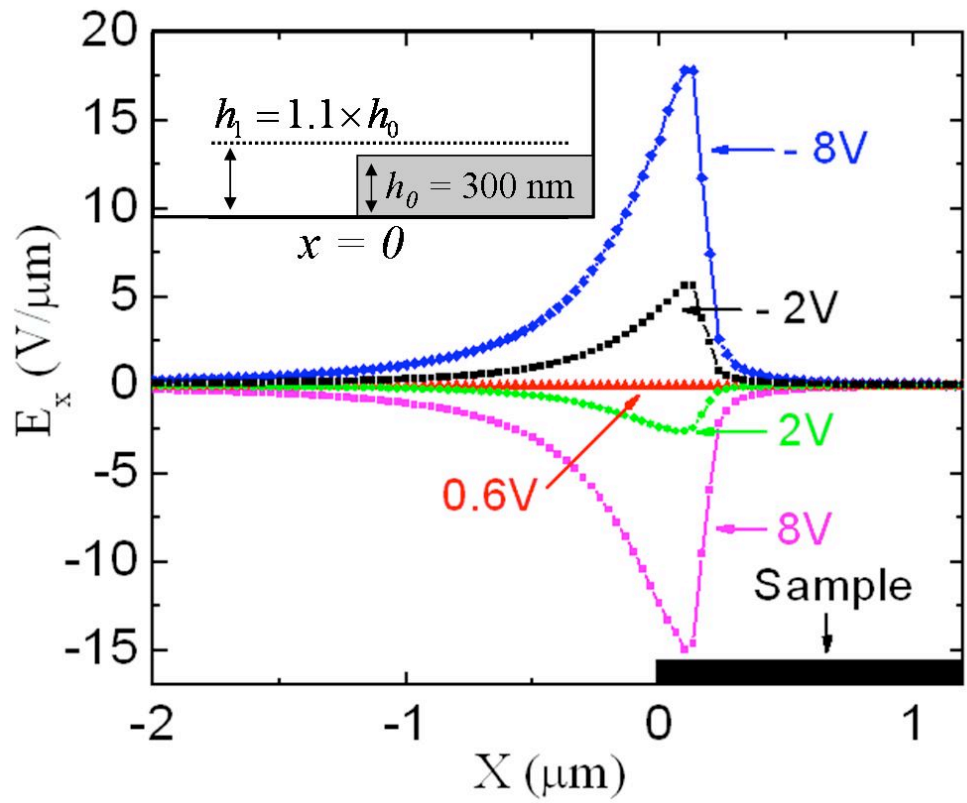

Figure 8. $E_x$ along the dotted line in inset at a height $h_1 = 1.1 \times h_0$ (330 nm) is plotted as a function of $x$. Magnitude of $E_x$ is zero for all $x$ when the sample is biased at 0.6 V.